\documentclass[aps,superscriptaddress,twocolumn,a4paper,showpacs]{revtex4}
\usepackage[colorlinks=true, pdfstartview=FitV, linkcolor=blue, citecolor=red, urlcolor=black]{hyperref}
\usepackage{graphicx}
\usepackage[all]{xy}
\usepackage{amsmath}
\usepackage{amssymb}
\usepackage{tensor}
\newcommand{\be}{\begin{equation}}
\newcommand{\ee}{\end{equation}}
\newcommand{\ben}{\begin{eqnarray}}
\newcommand{\een}{\end{eqnarray}}
\newcommand{\bes}{\begin{subequations}}
\newcommand{\ees}{\end{subequations}}
\def\bal#1\eal{\begin{align}#1\end{align}}

\newcommand{\LL}{{\cal L}}

\newcommand{\vphi}{{\varphi}}

\begin{document}

\title{Long range vortex configurations in generalized models \\ with the Maxwell or Chern-Simons dynamics}
\author{I. Andrade}\affiliation{Departamento de F\'\i sica, Universidade Federal da Para\'\i ba, 58051-970 Jo\~ao Pessoa, PB, Brazil}
\author{D. Bazeia}\affiliation{Departamento de F\'\i sica, Universidade Federal da Para\'\i ba, 58051-970 Jo\~ao Pessoa, PB, Brazil}
\author{M.A. Marques}\affiliation{Departamento de F\'\i sica, Universidade Federal da Para\'\i ba, 58051-970 Jo\~ao Pessoa, PB, Brazil}
\author{R. Menezes}\affiliation{Departamento de Ci\^encias Exatas, Universidade Federal
da Para\'{\i}ba, 58297-000 Rio Tinto, PB, Brazil}
\affiliation{Departamento de F\'\i sica, Universidade Federal da Para\'\i ba, 58051-970 Jo\~ao Pessoa, PB, Brazil}

\date{\today}

\begin{abstract}
In this work we deal with vortices in Maxwell-Higgs or Chern-Simons-Higgs models that engender long range tails. We find first order differential equations that support minimum energy solutions which solve the equations of motion. In the Maxwell scenario, we work with generalised magnetic permeabilities that lead to vortices described by solutions, magnetic field and energy density with power-law tails that extend farther than the standard exponential ones. We also find a manner to obtain a Chern-Simons model with the same scalar and magnetic field profiles of the Maxwell case. By doing so, we also find vortices with the aforementioned long range feature, which is also present in the electric field in the Chern-Simons model. The present results may motivate investigations on nonrelativistic models, in particular in the case involving Rydberg atoms, which are known to present long range interactions and relatively long lifetimes. 
\end{abstract}

\pacs{11.27.+d}
\maketitle
\section{Introduction}
In high energy physics, vortices are planar structures that appear under the action of a complex scalar field coupled to a gauge field under an $U(1)$ local symmetry \cite{manton,wein}. The first relativistic model investigated was the well-known Nielsen-Olesen one \cite{NO}, whose gauge field is controlled by the Maxwell term. In this case, the vortex is electrically neutral and engenders quantized flux. The equations of motion that control the fields are of second order. To simplify the problem, it was shown in Ref.~\cite{bogopaper} that, by using arguments of minimal energy, one can find first order equations that are compatible with the equations of motion. Even though the analytical form of the solutions remain unknown in terms of known functions, one can estimate their behavior out of their core, which is asymptotically dominated by an exponential function.

A distinct possibility to investigate vortices is by exchanging the Maxwell term with the Chern-Simons one, as firstly investigated by Jackiw and Weinberg \cite{jackiw}, and by Hong, Kim and Pac \cite{coreanos}. In this scenario, the vortex becomes electrically charged, with quantized charge. A first order formalism may also be developed here and, as in the Maxwell case, only the numerical solutions are found. In a way similar to the Maxwell-Higgs vortices, one can show that the Chern-Simons vortices present asymptotic behavior that are also ruled by an exponential behavior.

In the above standard models, the first order formalism requires the potential to engender a fourth-order power in the scalar field for the Maxwell case in Ref.~\cite{bogopaper} and a sixth-order power for the Chern-Simons model described in Refs.~\cite{jackiw,coreanos}. This means that one does not have the freedom to choose a potential that leads to distinct features. A possibility to circumvent this issue is by including extra functions that depend on the scalar field, additionally to the potential. For instance, in the Maxwell model, one may consider a generalized magnetic permeability. In the Chern-Simons scenario, the magnetic permeability cannot be modified, since it would break gauge invariance, so one can make use of a function that drives the dynamical term of the scalar field. Over the years, several papers dealing with vortices in generalized models appeared in the literature making use of other types of generalizations, such as the Born-Infeld dynamics and powers of the dynamical term of the scalar field; see, e.g., Refs.~\cite{genv1,genv2,genv3,genv4,genv5,genv6,genv7,genv8,compvortex,compcs,twin,godvortex}. This brings to light distinct features, such as uniform magnetic field inside the structure, compact vortices and the existence of twinlike models, which are models that support the very same localized solution with the same energy density.

In the study of kinks in $(1,1)$ dimensions, the standard solutions such as the ones of $\phi^4$ and sine-Gordon models engender exponential tail. For potentials with null classical mass at the minima, the asymptotic behavior is controlled by polynomial functions; see Refs.~\cite{long1,long2,long3,long4,long5,long6}. Since the tail of the structure extends farther than the ones of the standard case, they are called long range kinks. Long range structures may also arise in the study of non topological solitons, whose standard model only support power law tails \cite{longv1,longv2,longv3}. A similar behavior also arises in the study of both topological and non topological vortices in models with non minimal coupling \cite{longv4}.

In this work, we seek for vortex configurations that exhibit long range tails in both Maxwell-Higgs and Chern-Simons-Higgs scenarios. We first consider the Maxwell-Higgs model in Sec. \ref{MHM} and then, by using a procedure that we will introduce in Sec.~\ref{cssec}, we show how to obtain a Chern-Simons-Higgs model that support the same scalar and magnetic field configurations of a Maxwell-Higgs one. To illustrate the method, we take a model that support analytical solutions with polynomial tails in the Maxwell-Higgs scenario, found in Ref.~\cite{ana1}. In this case, the Chern-Simons model with the same scalar and magnetic fields requires the addition of awkward functions in the Lagrange density, so we also include a novel model that engender the long range behavior in both scenarios. We conclude the work in Sec. \ref{E}.

Before starting the investigation, we emphasize that the presence of vortices with long range tails in high energy physics may trigger further interest on this kind of configuration, since the distinct tail may ultimately modify the way they interact with one another, leading to a novel collective behavior. This is the main motivation of this work, and we think it can also attract interest to nonrelativistic models, in particular to the case of the Gross-Pitaevskii equation, which is appropriate to describe vortex excitations in Bose-Einstein condensates \cite{conden,conden2}. An interesting possibility relies on the use of Rydberg atoms, which engender very large principal quantum numbers, long range interactions and relatively long lifetimes \cite{Ry,Detect}. Another possibility concerns the study of cold and ultracold hybrid ion-atom  systems \cite{R7}. 

\section{Maxwell-Higgs Model}\label{MHM}
We consider a gauge field and a complex scalar field coupled through an $U(1)$ local symmetry in $(2,1)$ flat spacetime dimensions, with metric $\eta_{\alpha\beta}={\rm diag}(+,-,-)$ and action $S=\int d^3x\LL$, where the Lagrange density is taken with dimensionless fields and coordinates, in the form
\be\label{lmodel}
\LL= -\frac{1}{4\mu(|\vphi|)} F_{\alpha\beta}F^{\alpha\beta} + \overline{D_\alpha \vphi}D^\alpha \varphi   - V(|\vphi|).
\ee
Here, we have $D_\alpha = \partial_\alpha +iA_\alpha$, $F_{\alpha\beta}=\partial_\alpha A_\beta - \partial_\beta A_\alpha$ and the overline stands for complex conjugation. In this case, $\mu(|\vphi|)$ denotes a generalized magnetic permeability. The equations of motion of the fields $\vphi$ and $A_\alpha$ associated to the Lagrange density \eqref{lmodel} are
\bes\label{geom}
\begin{align}
& D_\alpha D^\alpha \vphi + \frac{\vphi}{2|\vphi|}\!\left(-\frac{\mu_{|\vphi|}}{4\mu^2} F_{\alpha\beta}F^{\alpha\beta} + V_{|\vphi|} \right)\! =0,\\ \label{meqs}
& \partial_\alpha \left(\frac1\mu F^{\alpha\beta}\right) + 2\Im(\overline{\vphi}D^\beta\vphi)=0,
\end{align}
\ees
in which $\Im(z)$ represents the imaginary part of $z$, and we have used the notation $\mu_{|\vphi|}=d\mu/d|\vphi|$ and $V_{|\vphi|} = \partial V/\partial{|\vphi|}$. Invariance of the Lagrange density \eqref{lmodel} under spacetime translations leads to the energy-momentum tensor
\be\label{emt}
	T_{\alpha\beta}=\frac{1}{\mu}F_{\alpha\lambda}\tensor{F}{^\lambda_\beta} +  2\Re\left( \overline{D_\alpha \vphi}D_\beta \vphi\right) - \eta_{\alpha\beta} \LL,
\ee
where $\Re(z)$ denotes the real part of $z$. In the case of static configurations, we take $A_0=0$ knowing that the Gauss' law for our model, given by the temporal component of Eq.~\eqref{meqs}, is compatible with this condition. This makes the vortex being electrically neutral. We proceed the investigation by taking
\be\label{ansatz}
\vphi = g(r)e^{in\theta} \quad\text{and}\quad \textbf{A} = {\frac{\hat{\theta}}{r}\left(n-a(r)\right)},
\ee
where $r$ and $\theta$ are polar coordinates and $n$ is the vortex winding number. The functions $g(r)$ and $a(r)$ are monotonic and must obey the boundary conditions
\be\label{bcond}
g(0) = 0, \quad a(0)= n, \quad \lim_{r\to\infty} g(r) = 1, \quad \lim_{r\to\infty} a(r) = 0.
\ee
With this, the terms associated to the dynamics of each field become 
\bes\label{XYm}
\begin{align}
	\overline{D_\alpha \vphi}D^\alpha \varphi &=-{g^\prime}^2- \frac{a^2g^2}{r^2} \\
	-\frac{1}{4} F_{\alpha\beta}F^{\alpha\beta} &=-\frac{{a^\prime}^2}{2r^2},
\end{align}
\ees
where the prime denotes the derivative with respect to $r$. Furthermore, the magnetic field takes the form 
\be\label{b}
B=-a^\prime/r
\ee
and the magnetic flux $\Phi=2\pi\int_0^\infty r dr B(r)$ is quantized:
\be\label{mflux}
\Phi=2\pi n.
\ee
 The equations of motion \eqref{geom} with \eqref{ansatz} become
\bes\label{secansatz}
\begin{align}\label{secansatzg}
&\frac{1}{r} \left(r g^\prime\right)^\prime -\frac{a^2g}{r^2} = -\frac{\mu_g{a^\prime}^2}{4\mu^2r^2} + \frac12 V_g, \\\label{secansatza}
& r\left(\frac{a^\prime}{\mu r} \right)^\prime = 2 ag^2.
\end{align}
\ees
The energy density $\rho\equiv T_{00}$ can be calculated from the energy-momentum tensor \eqref{emt} and Eq.~\eqref{ansatz}; it takes the form
\be\label{rho}
\rho = \frac{{a^\prime}^2}{2\mu r^2} + {g^\prime}^2 +\frac{a^2 g^2}{r^2} +V(g),
\ee
where $a(r)$ and $g(r)$ are the solutions of the equations of motion \eqref{secansatz}. These solutions, however, are not easy to be obtained, since one must solve second order differential equations that are coupled with one another. To simplify the problem, we make use of the first order formalism developed in Ref.~\cite{godvortex}, which appears for the stressless condition, $T_{ij}=0$. In this case, we get the first order equations
 \be\label{fo}
 g^\prime = \pm \frac{ag}{r}\quad\text{and}\quad -\frac{a^\prime}{r} = \pm\mu(g)\left(1-g^2\right).
 \ee
The pair of equations for the upper and lower signs are related by $a\to-a$. Here, the potential must be written as
\be\label{pot}
V(g) = \frac{\mu(g)}{2}\left(1-g^2\right)^2
\ee
to ensure the first order equations \eqref{fo} are compatible with the equations of motion \eqref{secansatz}. We may also take advantage of this formalism to use an auxiliar function $W(a,g)$ such that the energy density in Eq.~\eqref{rho} can be expressed in terms of a total derivative, as
\be\label{rhow}
\rho = \frac1r\frac{dW}{dr}, \quad\text{with}\quad W(a,g) = -a\left(1-g^2\right).
\ee
After integrating the above energy density, we get $E=2\pi|n|$, which is the same for the standard Nielsen-Olesen vortex \cite{NO}. For simplicity, from now on we only consider unit vorticity, $n=1$. Thus, one must use the positive sign in the first order equations \eqref{fo}.

In this paper we are interested in find vortices with polynomial tails, which we call long range vortices. However, before going further, we review the asymptotic behavior of the standard vortex $(\mu=1)$, which is described by the first order equations
\be
 g^\prime = \frac{ag}{r}\quad\text{and}\quad -\frac{a^\prime}{r} = 1-g^2.
\ee
To see how the solutions behave far from the origin, we look at the boundary conditions \eqref{bcond} and write $a(r)=0+a_{asy}(r)$ and $g(r) = 1-g_{asy}(r)$. By substituting these functions in the above first order equations and linearizing them, one can show that
\be\label{asystdm}
a_{asy} \approx \lambda\,\sqrt{2\,r}\,e^{-\sqrt{2}\,r}\quad\text{and}\quad g_{asy} \approx \lambda\,\frac{e^{-\sqrt{2}\,r}}{\sqrt{r}},
\ee
where $\lambda$ is a constant that can be adjusted to fit the numerical simulations. We then see these expressions rapidly vanishes as $r$ increases due to the exponential factor. The generalized magnetic permeability, however, has allowed for the presence of different vortex configurations, such as the compact vortices that we found in Ref.~\cite{compvortex}. 

Since we are interested in long range vortices, we first reproduce the analytical solutions found in Refs.~\cite{ana1,ana2}, in our model \eqref{lmodel} with the magnetic permeability given by 
\be\label{plong}
\mu(g)=\frac{2sg^{2s-2}\left|1-g^{2s}\right|^{1+\frac1s}}{\left|1-g^2\right|},
\ee
where $s$ is a real parameter such that $s\geq1$. In this case, the potential in Eq.~\eqref{pot} has the form
\be\label{pot1}
V(g) = s g^{2s-2}|1-g^2|\,|1-g^{2s}|^{1+\frac{1}{s}}
\ee
and we must solve the equations in Eq.~\eqref{fo}, which become
\be\label{folong}
 g^\prime = \frac{ag}{r}\quad\text{and}\quad -\frac{a^\prime}{r} = 2sg^{2s-2}\left(1-g^{2s}\right)^{1+\frac1s}.
 \ee
It support the analytical solutions
\be\label{longsols}
a(r)=\frac{1}{1+r^{2s}}\quad\text{and}\quad g(r) = \frac{r}{\left(1+r^{2s}\right)^{\frac{1}{2s}}}.
\ee
Notice that the tail of these solutions is controlled by $a(r)\propto r^{-2s}$ and $1-g(r)\propto r^{-2s}$, which is a distinct behavior from the exponential one found in Eq.~\eqref{asystdm}. The polynomial tail goes slower than the standard one to the boundary value. This shows the longe range character of the vortex. The magnetic field \eqref{b} and energy density \eqref{rhow} are
\bes
\bal
B(r) &= \frac{2sr^{2s-2}}{\left(1+r^{2s}\right)^2}, \\ \label{rho1m}
\rho(r) &=\frac{2\left(1 -sr^{2s} +sr^{2s-2}\left(1+r^{2s}\right)^{\frac1s}\right)}{\left(1+r^{2s}\right)^{2+\frac1s}}.
\eal
\ees
They can be integrated to give flux $\Phi=2\pi$ and energy $E=2\pi$. Notice that both $E$ and $\Phi$ do not depend on $s$, as previously informed.

\begin{figure}[t!]
\centering
\includegraphics[width=8.2cm,trim={0.6cm 1cm 0 0},clip]{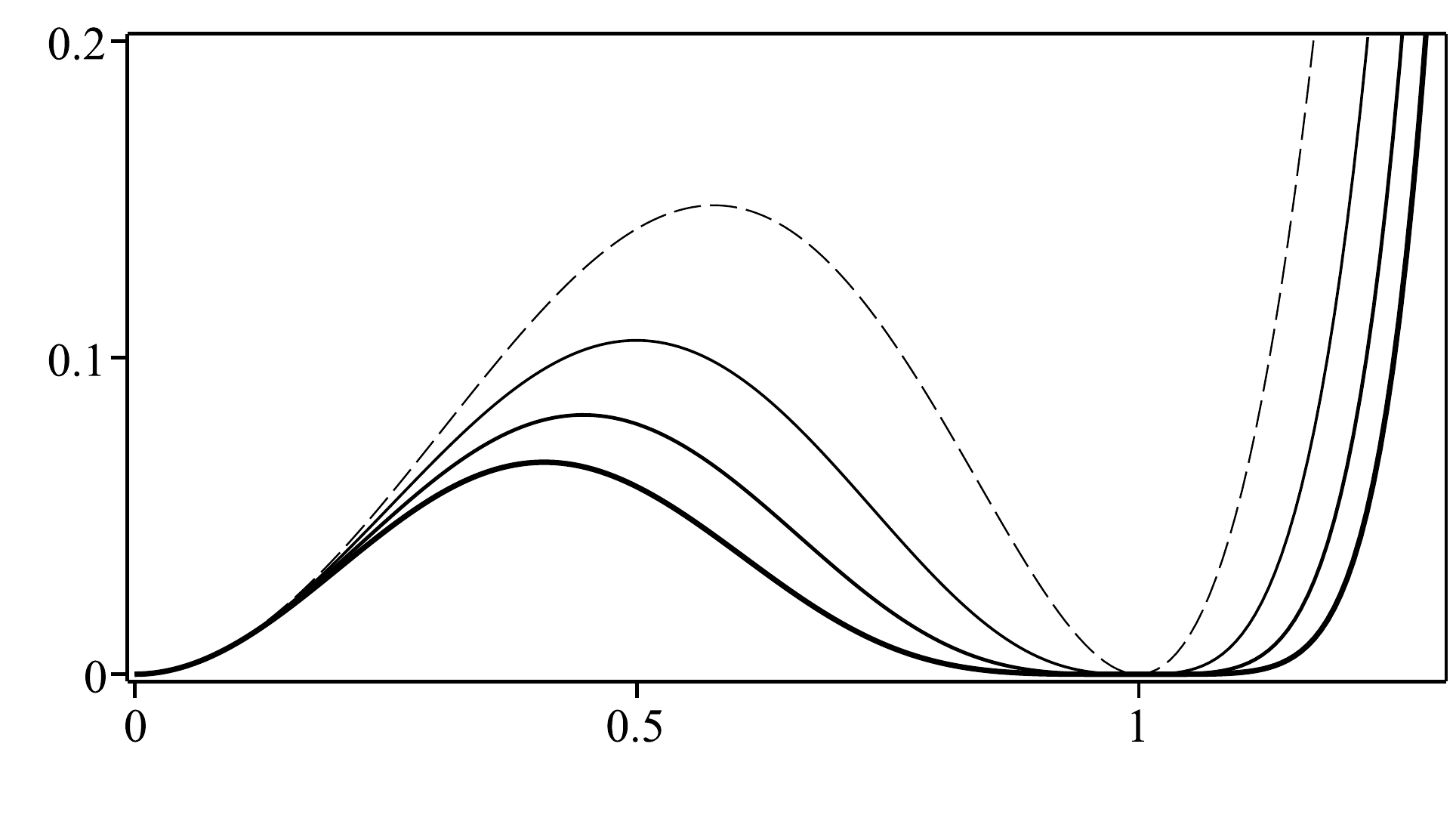}
\caption{The potential in Eq.~\eqref{potnum} as a function of $g$, $V(g)$, for $l=1,2,3$ and $4$. The dashed line represents the case $l=1$ and the thickness of the lines increases with $l$.}
\label{figpotnum}
\end{figure}

We now introduce a novel model that supports vortex configurations with long range tails. It is given by the magnetic permeability
\be\label{munum}
\mu(g) = 2g^2\left|1-g^2\right|^{l-1},
\ee
where $l$ is a real parameter such that $l\geq1$. The case $l=1$ recovers the model investigated in Ref.~\cite{leenam}, which reproduces, using a generalized magnetic permeability, the standard Chern-Simons solutions $a(r)$ and $g(r)$ \cite{jackiw,coreanos}, whose tails are dominated by an exponential function, similarly to the behavior in Eq.~\eqref{asystdm}. So, for a general $l$, the potential in Eq.~\eqref{pot} becomes
\be\label{potnum}
V(g) = g^2\left|1-g^2\right|^{l+1}.
\ee
This potential is displayed in Fig.~\ref{figpotnum} for some values of $l$. One can show that $d^mV/dg^m|_{g=1}=0$ for $m=0,\ldots,\left\lceil{l}\right\rceil$, where $\left\lceil{l}\right\rceil$ denotes the ceiling function.

In this model, the first order equations \eqref{fo} take the form
\be\label{eqnumm}
 g^\prime = \frac{ag}{r}\quad\text{and}\quad -\frac{a^\prime}{r} = 2g^2\left(1-g^2\right)^l.
\ee
\begin{figure}[t!]
\centering
\includegraphics[width=8.2cm,trim={0.6cm 1cm 0 0},clip]{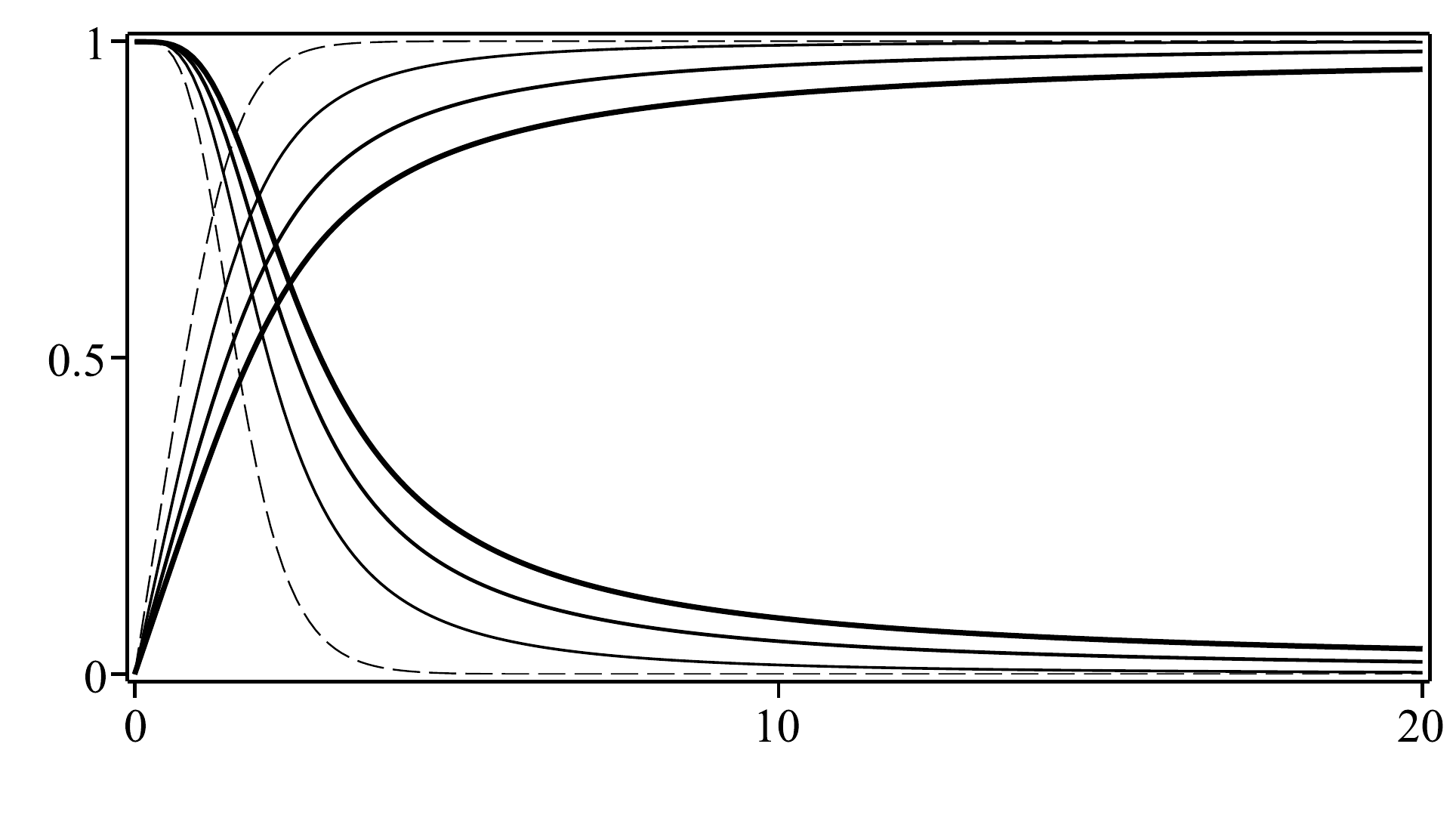}
\includegraphics[width=8.2cm,trim={0.6cm 1cm 0 0},clip]{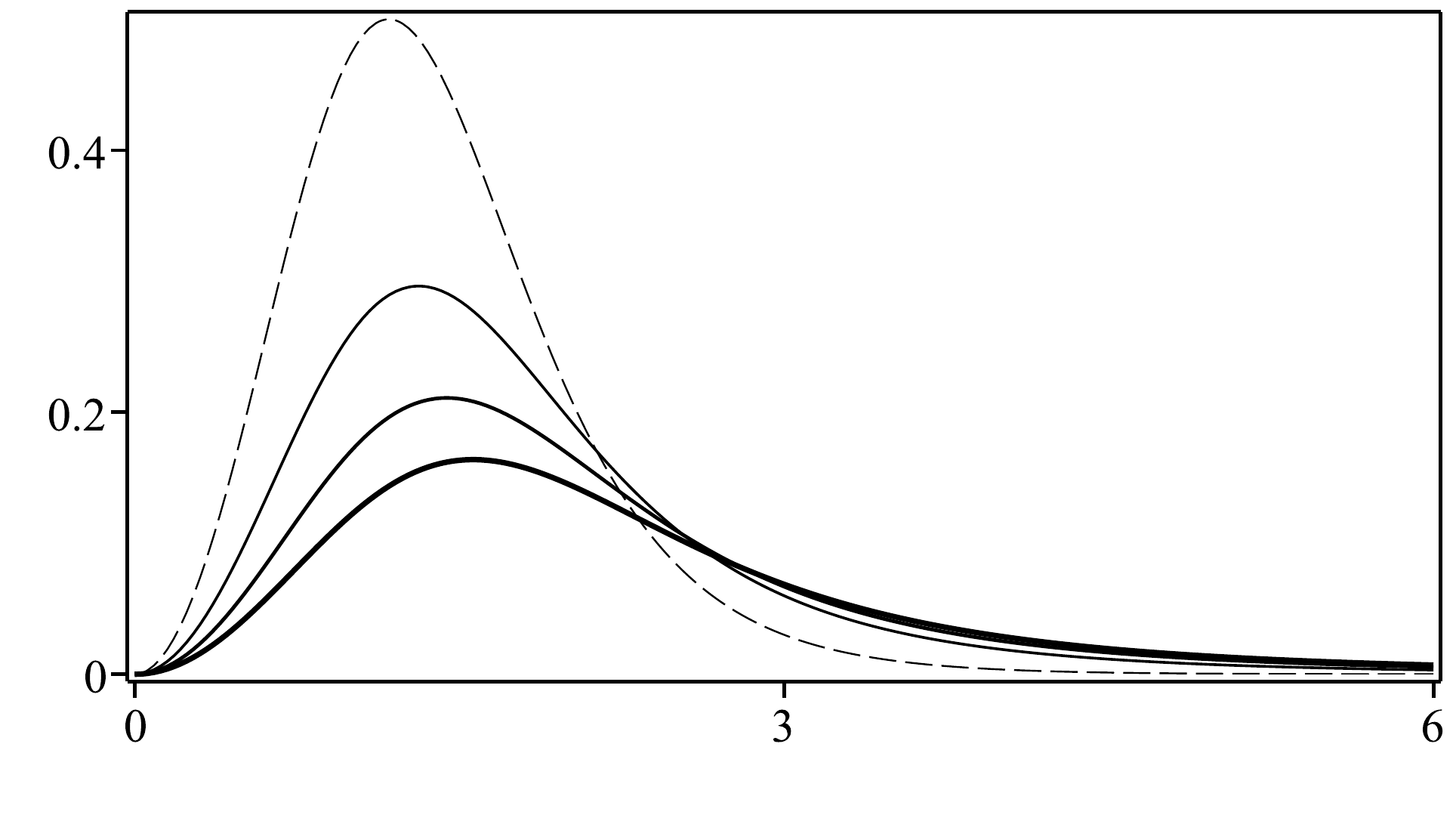}
\caption{The solutions (top) $a(r)$ (descending lines) and $g(r)$ (ascending lines) of Eq.~\eqref{eqnumm} and the magnetic field $B(r)$ (bottom) for $l=1,2,3$ and $4$. The dashed lines represent the case $l=1$ and the thickness of the lines increases with $l$.}
\label{figsolbnum}
\end{figure}
The above first order equations admit the asymptotic behavior
\bes\label{solasynum}
\begin{align}
	a_{asy}(r) &\approx (l-1)^{-\frac{l+1}{l-1}}\, r^{-\frac{2}{l-1}},\\
	g_{asy}(r) &\approx \frac{1}{2}\,(l-1)^{-\frac{2}{l-1}}\, r^{-\frac{2}{l-1}}.
\end{align}
\ees
The above expressions show that the solutions exhibit a polynomial tail that goes slower to their boundary values as $l$ increases. One may also verify that the magnetic field \eqref{b} and the energy density \eqref{rhow} behaves asymptotically as
\bes\begin{align}
B(r) &\approx2\left(l-1\right)^{-\frac{2l}{l-1}}\,r^{-\frac{2l}{l-1}}\\
\rho(r) &\approx 4\left(l-1\right)^{-\frac{2(l+1)}{l-1}}\,r^\frac{-2(l+1)}{l-1}.
\end{align}
\ees
Thus, similarly to the solutions, both the magnetic field and the energy density engender polynomial tails. As in the previous model, the quantities that describe the vortex present a power-law asymptotic behavior which shows the long range behavior of the structure.

\begin{figure}[t!]
\centering
\includegraphics[width=8.2cm,trim={0.6cm 1cm 0 0},clip]{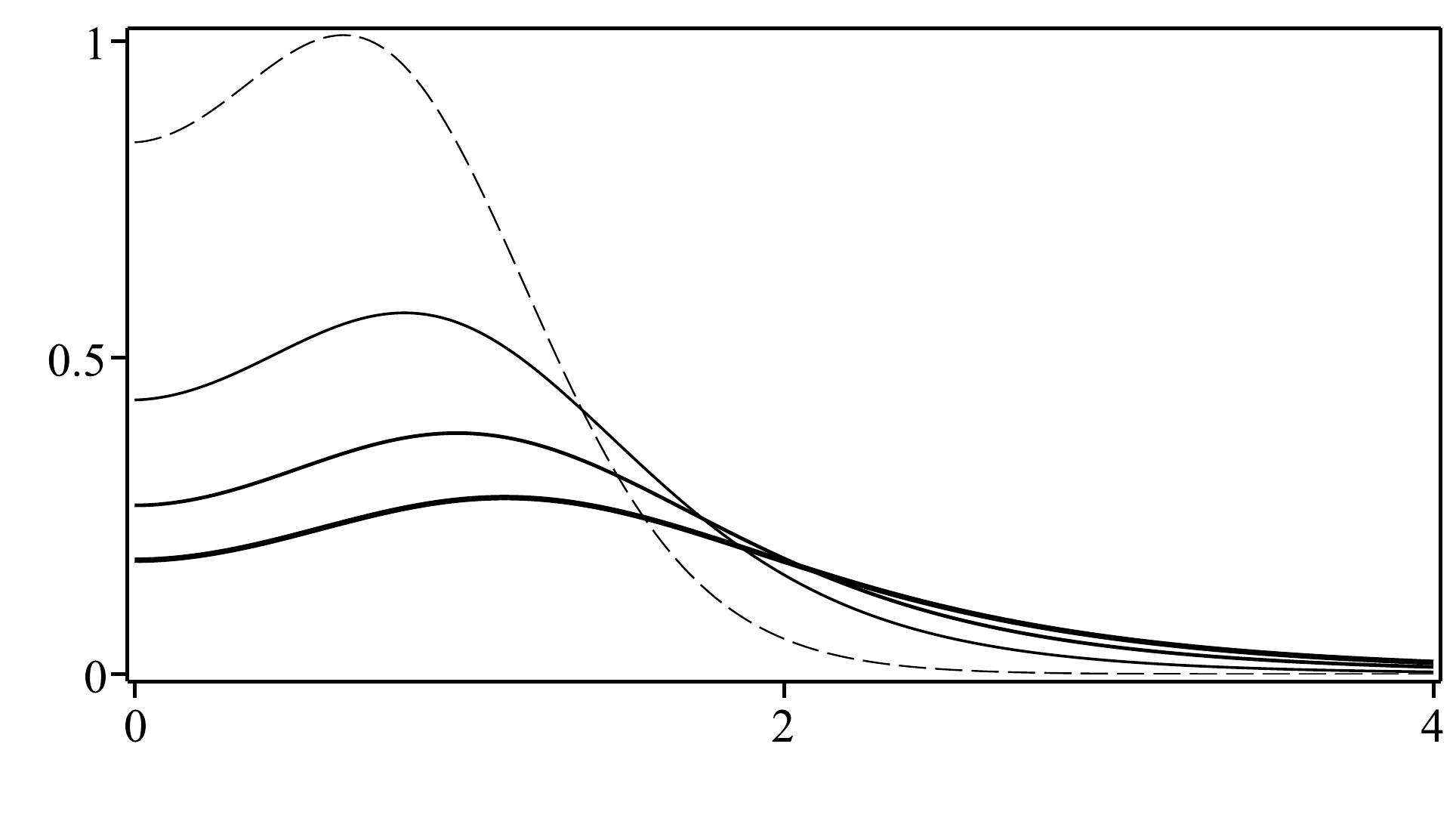}
\caption{The energy density $\rho(r)$ associated to the solutions of Eq.~\eqref{eqnumm} for $l=1,2,3$ and $4$. The dashed line represents the case $l=1$ and the thickness of the lines increases with $l$.}
\label{figrhonum}
\end{figure}

Differently from the previous model, here we were not able to find the analytical solutions of the first order equations \eqref{eqnumm}. So, we must use numerical procedures to solve them for each $l$. In Fig.~\ref{figsolbnum}, we display the solutions and the magnetic field $B(r)$ for some values of $l$. We also calculate the energy density numerically and show it in Fig.~\ref{figrhonum}. One can see that, for $l>1$, all the quantities that describe the vortex configuration take larger distances to attain their boundary values when compared to the standard case with exponential tails. Moreover, as $l$ increases, the tails gets larger and larger.
\section{Chern-Simons Models}\label{cssec}
We now exchange the Maxwell term for the Chern-Simons one in the Lagrange density \eqref{lmodel}. In this case, we cannot include a factor depending on the scalar field multiplying the Chern-Simons term, because it would break gauge invariance. Nevertheless, as we have shown in Ref.~\cite{godvortex}, we need generalized models to find vortices with features that differ from the standard ones \cite{jackiw,coreanos}. Then, we consider the generalized class introduced in Ref.~\cite{kcs}, which includes a factor that modifies the dynamical term of the scalar field
\begin{equation}\label{lcs}
\LL = \frac{1}{4}\epsilon^{\alpha\beta\gamma}A_\alpha F_{\beta\gamma} + K(|\vphi|)\overline{D_\alpha \vphi}D^\alpha\vphi -V(|\vphi|).
\end{equation}
The equations of motion associated to the above Lagrange density are
\bes\label{eomcs}
\begin{align}
& D_\alpha (K D^\alpha\vphi)= \frac{\vphi}{2|\vphi|}\left(K_{|\vphi|}\overline{D_\alpha \vphi}D^\alpha\vphi -V_{|\vphi|} \right), \\
 &\frac{\kappa}{2} \epsilon^{\alpha\beta\gamma}F_{\alpha\beta} + 2K\,2\Im(\overline{\vphi}D^\gamma\vphi)=0,
\end{align}
\ees
The energy momentum tensor has the form
\be\label{emtcs}
\begin{aligned}
T_{\alpha\beta}&=2K(|\vphi|)\Re\left( \overline{D_\alpha \vphi}D_\beta \vphi\right)\\
&\hspace{4mm}- \eta_{\alpha\beta} \left( K(|\vphi|)\overline{D_\lambda \vphi}D^\lambda\vphi -V(|\vphi|) \right).
\end{aligned}
\ee

Here, we cannot take $A_0=0$ as in the Maxwell case because this is not compatible with the equations of motion. So, we take the fields in the form of Eq.~\eqref{ansatz} with their usual boundary conditions, and $A_0=A_0(r)$. One can show that the magnetic field is $B=-a^\prime/r$ and its associated flux is given by Eq.~\eqref{mflux}. Here, we have an additional feature that arises due to the presence of the temporal component of the gauge field: the electric field, whose intensity is $|\textbf{E}|=|A_0^\prime|$. In this case, the vortex is electrically charged, with charge $Q=-\Phi$. For simplicity, we take unit vorticity, $n=1$. From Eqs.~\eqref{eomcs}, we get the following equations of motion
\bes\label{eomcsansatz}
\begin{align}
\frac{1}{r} \left(rK g^\prime\right)^\prime + K g \left(A_0^2-\frac{a^2}{r^2} \right)+& \nonumber\\
+ \frac12 \left(\left(g^2A_0^2-{g^\prime}^2-\frac{a^2g^2}{r^2}\right)K_{g} -V_{g} \right) &= 0, \\ \label{a0csansatz}
 \frac{a^\prime}{r} + 2K g^2 A_0 &= 0, \\ \label{ecsansatz}
 {A_0^\prime} + \frac{2K a g^2}{r} &= 0.
\end{align}
\ees
Also, the energy density is calculated from the component $T_{00}$ in Eq.~\eqref{emtcs} and takes the form
\be\label{rhoans}
\rho =\frac14 \frac{{a^\prime}^2}{r^2 g^2K(g)} + \left({g^\prime}^2+\frac{a^2g^2}{r^2}\right)K(g) + V(g).
\ee
The equations of motion \eqref{eomcsansatz} are of second order. To simplify the problem, we follow the first order formalism developed in Ref.~\cite{godvortex} to obtain
\be\label{focs1}
g^\prime = \frac{ag}{r} \quad\text{and}\quad -\frac{a^\prime}{r} =2g\sqrt{KV}.
\ee
The potential, however, cannot have an arbitrary form because the above equations must be compatible with the equations of motion \eqref{eomcsansatz}. One can show that the functions $K(g)$ and $V(g)$ are constrained to obey
\be\label{constraintVK}
\frac{d}{dg} \left(\sqrt{\frac{V}{g^2K}}\,\right) = -2gK.
\ee
The above equation allows us to write the potential as
\be
V(g) = 4g^2K(g)\left(\int dg\, gK(g)\right)^2,
\ee
in which an integration constant always arise in the process, since we are dealing with an indefinite integration. For a general $K(g)$, the first order equations \eqref{focs1} become
\bes\label{focs}
\bal
g^\prime &= \frac{ag}{r} \\ \label{foacs}
-\frac{a^\prime}{r} &= -4g^2K(g)\!\int\! dg \,gK(g)
\eal
\ees
such that one must choose $K(g)$ and the integration constant to get solutions compatible with the boundary conditions \eqref{bcond}. In this case, the energy density is given by
\be\label{wcs}
\rho = \frac1r\frac{dW}{dr},\quad\text{where}\quad W(a,g) = 2a\int\! dg \,gK(g).
\ee
By integrating this energy density, we get energy 
\be\label{energycs}
E=2\pi\left|W(a=0,g=1)-W(a=1,g=0)\right|.
\ee
So, the function $K(g)$ and the integration constant that arises in the process also modifies the energy of the vortex.

The simplest example is the standard case, $K(g)=1$, investigated in Refs.~\cite{jackiw,coreanos}. In this situation, to develop the Bogomol'nyi procedure \cite{bogopaper}, one must take the $|\vphi|^6$ potential, given by 
\be\label{potjackiw}
V(g) = g^2\!\left(v^2-g^2\right)^2,
\ee
where $v$ is the symmetry breaking parameter. To ensure that the first order equations \eqref{focs} support solutions compatible with the boundary conditions in Eq.~\eqref{bcond} we set $v=1$. By doing so, the model is governed by
\be\label{eqjackiw}
g^\prime=\frac{ag}{r} \quad\text{and}\quad -\frac{a^\prime}{r} = 2g^2(1-g^2).
\ee
As one knows, the analytical solutions of these equations remain unknown. The energy density is written as in Eq.~\eqref{wcs} with $W = a(g^2-1)$. So, from Eq.~\eqref{energycs}, one has the energy $E=2\pi$.

We note here that the first order equation for $a(r)$ in both Maxwell-Higgs and Chern-Simons-Higgs models has the form
\be\label{af}
-\frac{a^\prime}{r} = f(g),
\ee
which is always solved with $g^\prime=ag/r$. Notice that $B=-a^\prime/r$ only depends on $f(g)$; this allows us to find models in both scenarios with the same solutions and magnetic field. For $f(g)=\mu(g)(1-g^2)$, we get the Maxwell-Higgs model with the presence of the generalized magnetic permeability $\mu(g)$. On the other hand, comparing the above equation with the in Eq.~\eqref{foacs}, we see that for $f(g) = -4g^2K(g)\!\int\! dg \,2gK(g)$ we obtain the Chern-Simons case. This means that one may relate both models. In particular, for a known $f(g)$, one can show that the Chern-Simons-Higgs model is obtained through
\be\label{kf}
K(g) = \frac{f(g)}{2g^2}\left(-\int d(g^2)\, \frac{f(g)}{g^2}\right)^{-\frac12}.
\ee
One must be careful with this integration, because the integration constant must be properly chosen to make the above function be non negative in the interval where the solution $g(r)$ exists, i.e., $g\in[0,1]$, as stated in the boundary conditions \eqref{bcond}. Moreover, it must also lead to non-negative finite energy. In this case, the potential can be calculated from the right equation in \eqref{focs1} and Eq.~\eqref{af}; it is simply given by
\be\label{vf}
V(g) = \frac14\frac{f^2(g)}{g^2K(g)},
\ee
and the function $W(a,g)$ in Eq.~\eqref{wcs}, involved in the energy, is calculated in terms of $f(g)$ as
\be\label{wf}
W(a,g) = -a\,\frac{f(g)}{2g^2K(g)}.
\ee
Let us consider the standard case, $K(g)=1$, investigated in Ref.~\cite{jackiw,coreanos}. As we have commented before, in this case one gets the potential in Eq.~\eqref{potjackiw} with $v=1$ to match the boundary conditions \eqref{bcond}. We can substitute this in Eq.~\eqref{focs1} or use Eq.~\eqref{focs} to obtain the first order equations \eqref{eqjackiw}. Comparing this with Eq.~\eqref{af}, one can show that $f(g) = 2g^2(1-g^2)$. By using Eq.~\eqref{kf}, we get $K(g) = |1-g^2|/\sqrt{C-2g^2+g^4}$ and the potential $V(g) = g^2\left|1-g^2\right|\sqrt{C-2g^2+g^4}$. Notice there is an integration constant, $C$, in these expressions. Nevertheless, $K(g)$ has singularities for $C<1$, which we avoid here and the potential is V-shaped for $C>1$. So, we take $C=1$, which recovers the standard case, $K(g)=1$, and is the only choice that leads to a smooth potential.

Now, we use the procedure to obtain a Chern-Simons model that engender the same analytical solutions in Eq.~\eqref{longsols}. For $s=1$, the functions involved in the model lead to infinite energy, so we only consider $s>1$, for which
\bes\begin{align}
K(g) &= \sqrt{\frac{s(s-1)}{2}}\,\frac{g^{2s-4}\left|1-g^{2s}\right|^{1+\frac1s}}{H(g)},\\
V(g) &= \sqrt{\frac{2s^3}{s-1}}\,g^{2s-2}\left|1-g^{2s}\right|^{1+\frac1s}H(g),
\end{align}
\ees
with $H^2(g)=\left|1-g^{2s-2} \;_2F_1\!\left(-1-\frac1s,1-\frac1s;2-\frac1s;g^{2s}\right)\right|$, where $_2F_1(\alpha,\beta;\lambda;z)$ denotes the Hypergeometric function of parameters $\alpha,\beta$ and $\lambda$, and argument $z$. Also, we have chosen the integration constant $C=2s/(s-1)$ to obtain a simpler expression. The function $W(a,g)$ in the energy density that appears in Eq.~\eqref{wcs} is given by Eq.~\eqref{wf}, which leads to 
\be
\begin{aligned}
	W(a,g) &= -\sqrt{\frac{2s}{s-1}}\,a H(g).
\end{aligned}
\ee
So, for $s>1$, the energy is given by Eq.~\eqref{energycs} and has the form $E=2\pi\sqrt{2s/(s-1)}$. Notice that, even though the procedure works, it leads to exotic potentials, with the presence of a Hypergeometric function.

We carry on with the investigation and use the same method to get a Chern-Simons model that support the same solutions $a(r)$ and $g(r)$ of the Maxwell-Higgs model that we have introduced with the magnetic permeability \eqref{munum}. In this case, these solutions obey first order equations \eqref{eqnumm} and their associated magnetic field, all of them displayed in Fig.~\ref{figsolbnum}. Thus, we have $f(g) = 2g^2\left(1-g^2\right)^l$.

As we remarked below Eq.~\eqref{munum}, the case $l=1$ leads to solutions with exponential tails with a form similar to the one found in Eq.~\eqref{asystdm}, since both $a(r)$ and $g(r)$ are exactly the same of the standard Chern-Simons model \cite{jackiw,coreanos}, for any well-defined $K(g)$. Note however, that here, differently from the Maxwell-Higgs model described by the magnetic permeability \eqref{munum}, we have the presence of an electric field due to a nonvanishing temporal gauge component such that both of them depend on the form of the function $K(g)$. Since the purpose of our paper is to deal with long range vortices, we do not discuss the case $l=1$ with detail, using it only to compare the new solutions to the standard ones.

 To obtain the function $K(g)$, which controls the dynamical term of the scalar field in the Lagrange density, one must use Eq.~\eqref{kf}. It leads to
\be\label{kcsc}
K(g)= \sqrt{\frac{l+1}{2}}\,\frac{|1-g^2|^l}{\sqrt{\alpha+(1-g^2)^{l+1}}}.
\ee
Notice there is an integration constant $\alpha$ that appears in the process. It must be non-negative to ensure the above function is real. As we have shown in Eq.~\eqref{vf}, $K(g)$ determines the potential, which is given by
\be\label{potcsc}
V(g) = \sqrt{\frac{2}{l+1}}\,g^2\left|1-g^2\right|^l\sqrt{\alpha+(1-g^2)^{l+1}}.
\ee
The function $W(a,g)$ in Eq.~\eqref{wf} associated to the energy has the form
\be\label{wcsc}
W(a,g) = -\sqrt{2}\,a\,\sqrt{\frac{\alpha+\left(1-g^2\right)^{l+1}}{l+1}},
\ee
which makes the energy in Eq.~\eqref{energycs} be given by 
\be\label{energycsc}
E= 2\sqrt{2}\,\pi\,\sqrt{\frac{\alpha+1}{l+1}}.
\ee
The potential in Eq.~\eqref{potcsc} has a set of minima located at $g=1$ and at the origin, regardless the value of $\alpha$. Nevertheless, as we will show in this paper, the case $\alpha=0$ is special, so we deal with it later. First, we take $\alpha>0$. In this situation, we have $d^mV/dg^m|_{g=1}=0$ for $m=0,\ldots,\left\lceil{l-1}\right\rceil$.

As we have used the method in Eq.~\eqref{kf} to find the Chern-Simons model, the solutions $a(r)$ and $g(r)$, and the magnetic field are the same of the Maxwell case; see Fig.~\ref{figsolbnum}. However, we are now dealing with a vortex in the Chern-Simons scenario, so we also have the presence of $A_0$, which gives rise to an electric field, $\textbf{E}(r)$. It also modifies the energy density, which now depends on $K(g)$ as one can see in Eq.~\eqref{wcs}. Since we only know the numerical solutions, we estimate the asymptotic behavior of these quantities using the results for the tail of $a(r)$ and $g(r)$ in Eq.~\eqref{solasynum} substituted in Eqs.~\eqref{a0csansatz}, \eqref{ecsansatz} and \eqref{wcs}:
\bes\label{a0erhocsc}
\bal\label{a0csc}
A_0(r) &\approx \sqrt{\frac{2\alpha}{l+1}}\, \left(1+\frac{1}{2\alpha}\left(l-1\right)^{-\frac{2(l+1)}{l-1}}\,r^{-\frac{2(l+1)}{l-1}}\right)\\
|\textbf{E}(r)| &\approx \sqrt{\frac{2(l+1)}{\alpha}}\,\left(l-1\right)^{-\frac{3l+1}{l-1}}\,r^{-\frac{3l+1}{l-1}}\\
\rho(r) &\approx 2\,\sqrt{\frac{2\alpha}{l+1}}\,\left(l-1\right)^{-\frac{2l}{l-1}}\,r^{-\frac{2l}{l-1}}.
\eal
\ees
Thus, all of these quantities present a polynomial tail that is controlled by $l$, with $l\in(1,\infty)$. An interesting feature, is that $A_0$ tends to a non-null constant, such that $A_0\to \sqrt{2\alpha/(l+1)}$ for $r\to\infty$.

\begin{figure}[t!]
\centering
\includegraphics[width=8.2cm,trim={0.6cm 1cm 0 0},clip]{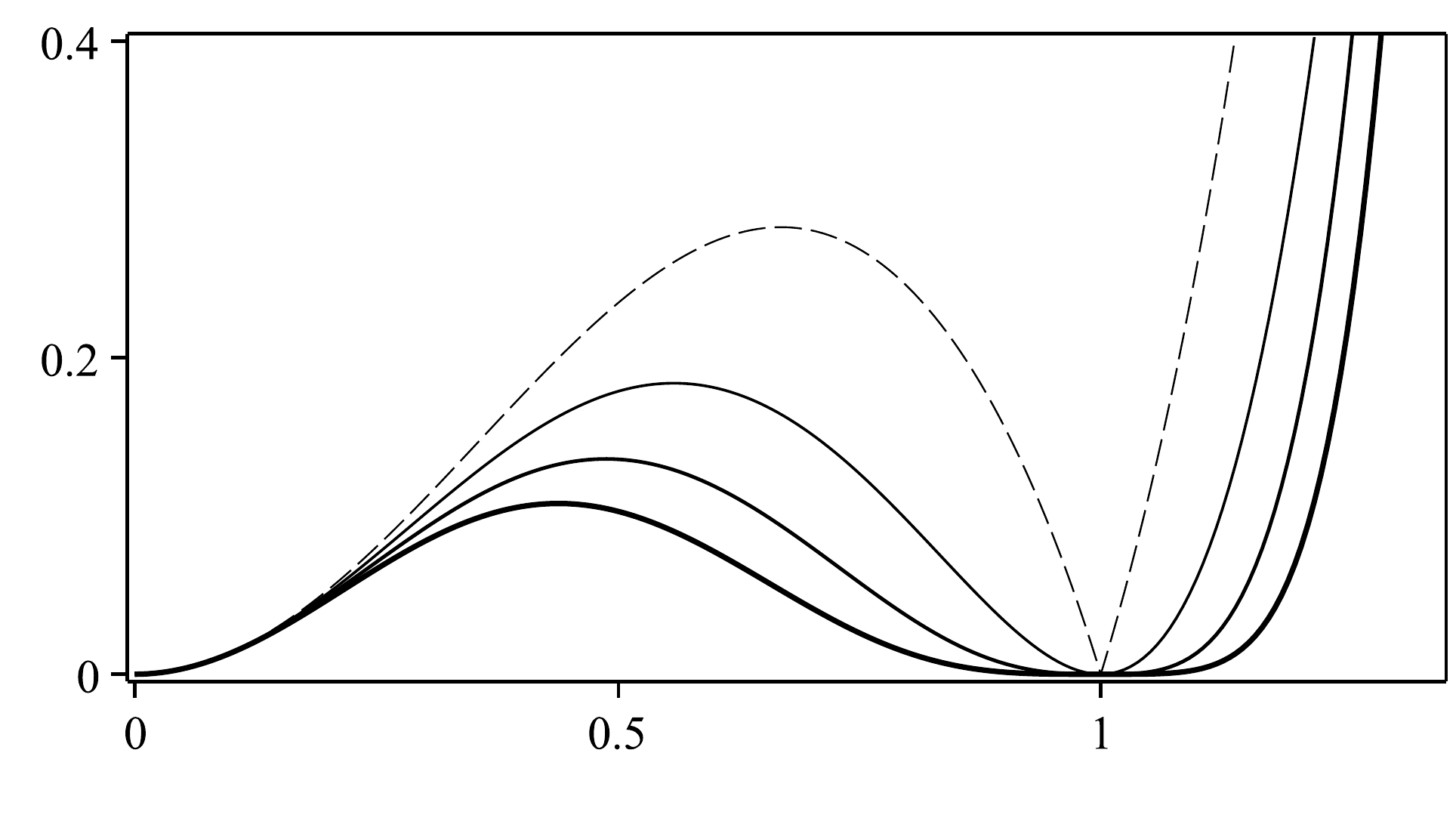}
\caption{The potential $V(g)$ in Eq.~\eqref{potcsc} with $\alpha=l$, for $l=1,2,3$ and $4$. The dashed line represents the case $l=1$ and the thickness of the lines increases with $l$.}
\label{figpotnumcsl}
\end{figure}

\begin{figure}[t!]
\centering
\includegraphics[width=8.2cm,trim={0.6cm 1cm 0 0},clip]{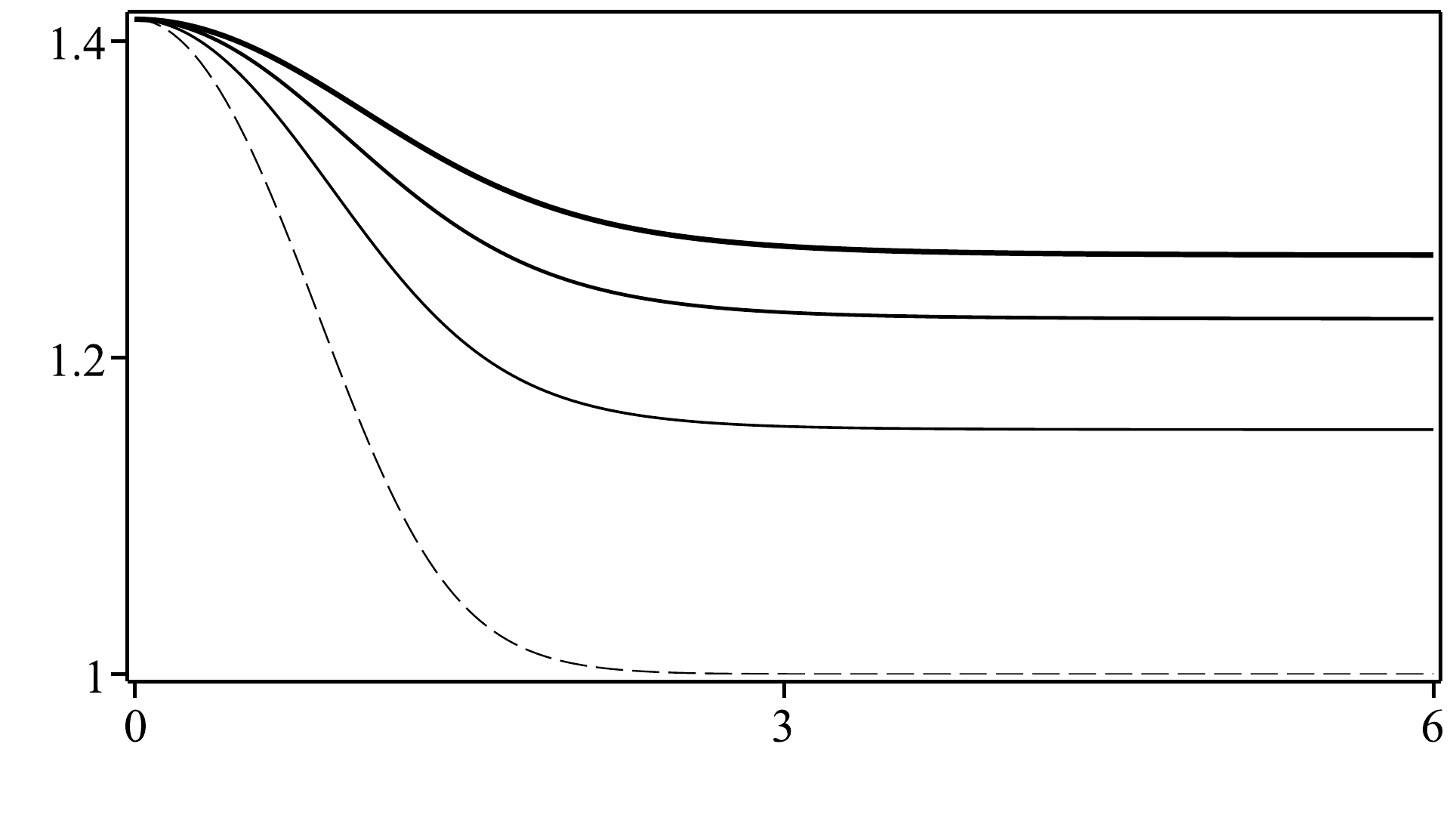}
\includegraphics[width=8.2cm,trim={0.6cm 1cm 0 0},clip]{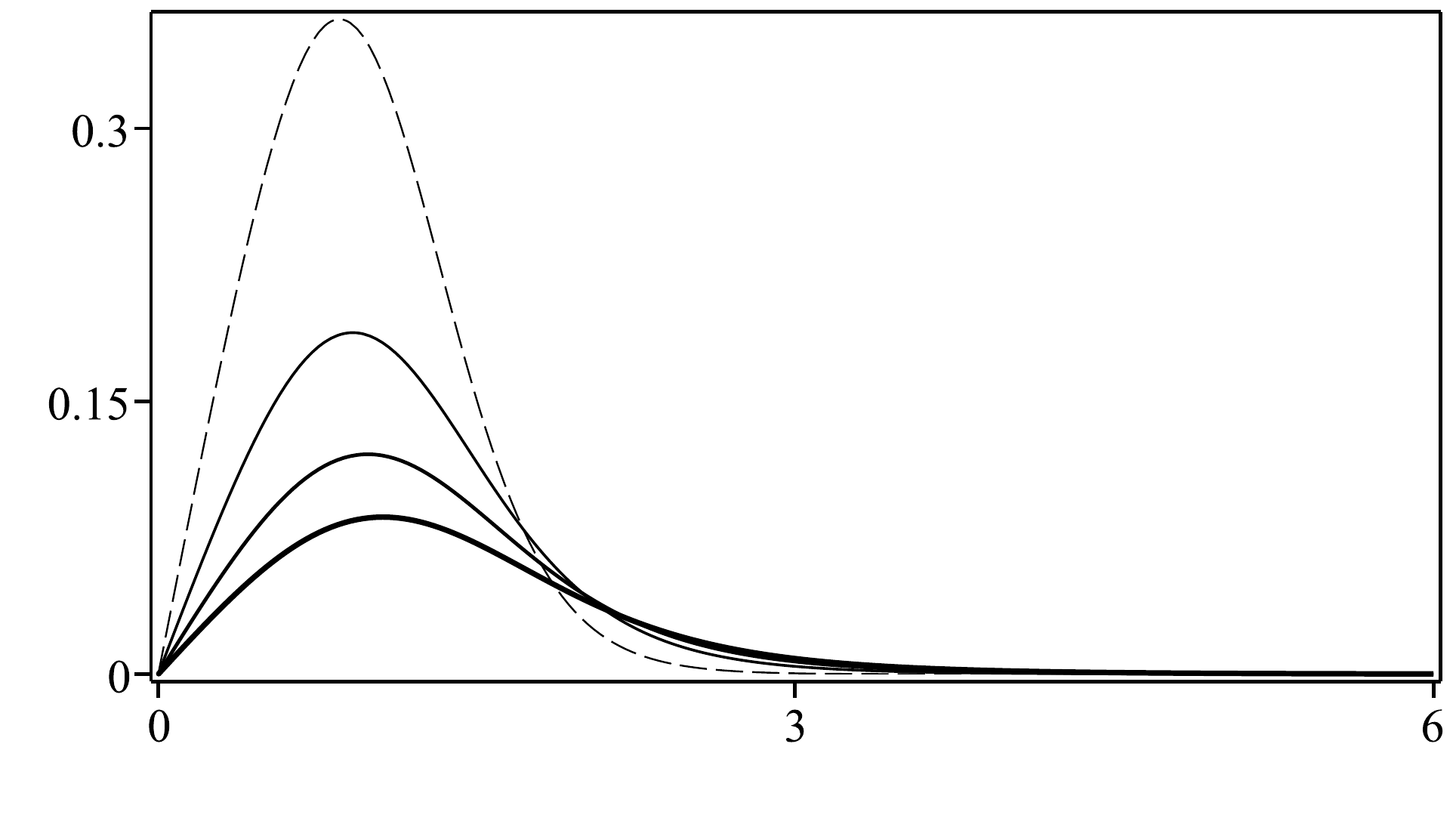}
\caption{The temporal component of the gauge field $A_0(r)$ (top) and the intensity of the electric field $|\textbf{E}(r)|$ (bottom) associated to the Chern-Simons model described by the potential in Eq.~\eqref{potcsc} with $\alpha=l$, for $l=1,2,3$ and $4$. The dashed lines represent the case $l=1$ and the thickness of the lines increases with $l$.}
\label{figenumcsl}
\end{figure}

\begin{figure}[t!]
\centering
\includegraphics[width=8.2cm,trim={0.6cm 1cm 0 0},clip]{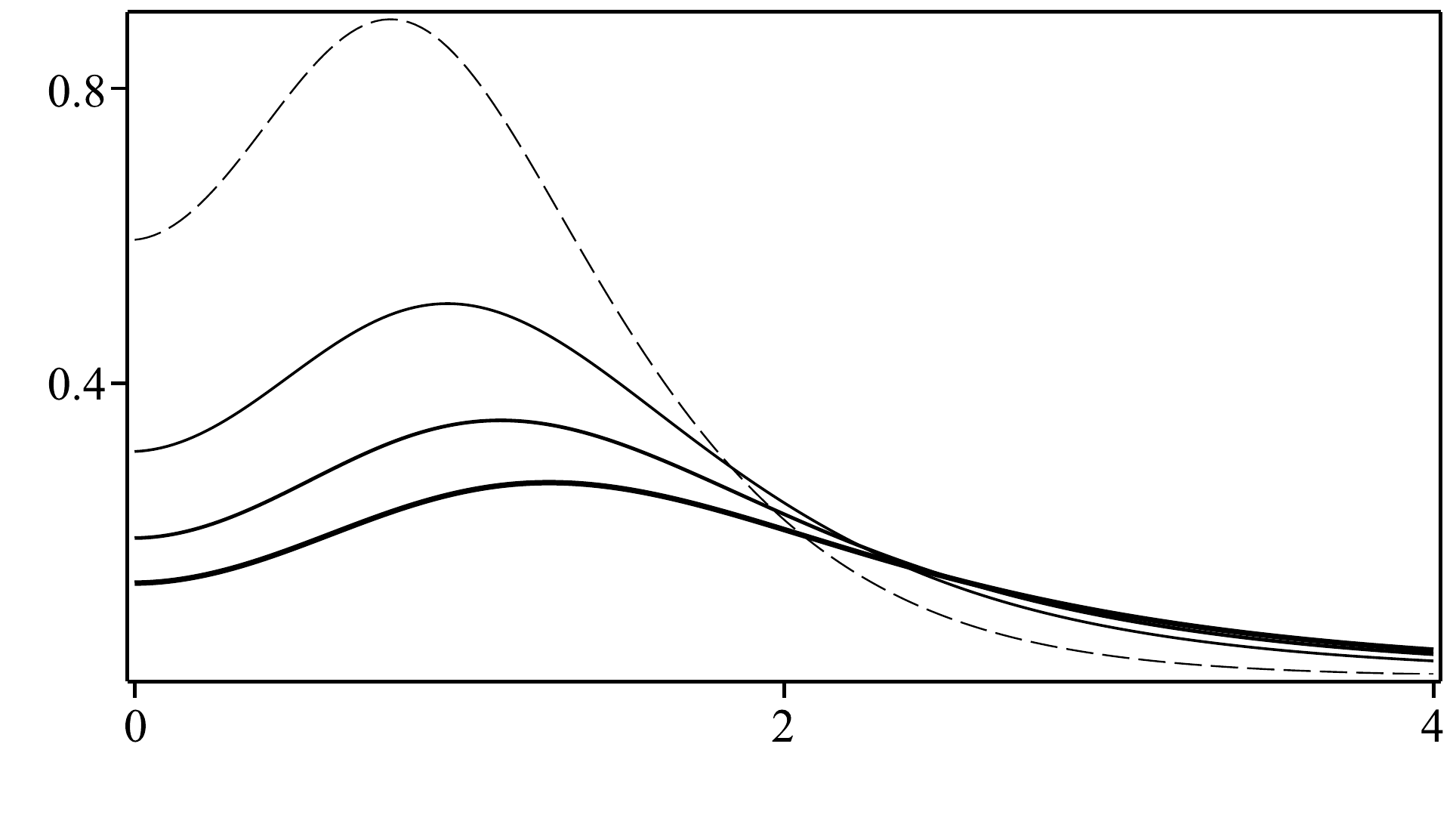}
\caption{The energy density $\rho(r)$ associated to the Chern-Simons model described by the potential in Eq.~\eqref{potcsc} with $\alpha=l$, for $l=1,2,3$ and $4$. The dashed lines represent the case $l=1$ and the thickness of the lines increases with $l$.}
\label{figrhonumcsl}
\end{figure}

An interesting case for positive $\alpha$ is $\alpha=l$, as it leads to vortices with fixed energy in Eq.~\eqref{energycsc}, $E=2\sqrt{2}\,\pi$, regardless the value of $l$. The function $K(g)$, the potential $V(g)$ and the other involved quantities can be calculated straightforwardly by taking $\alpha=l$ in Eqs.~\eqref{kcsc}-\eqref{a0erhocsc}. The potential can be seen in Fig.~\ref{figpotnumcsl} for some values of $l$. We then turn our attention to $A_0$, which gives rise to an electric field, $\textbf{E}(r)$. They can be calculated from Eqs.~\eqref{a0csansatz} and \eqref{ecsansatz} and are displayed in Fig.~\ref{figenumcsl}. From the graphic of $A_0$, one can see that, for $l\to\infty$ and $r\to\infty$, $A_0\to\sqrt{2}$. We also plot the energy density \eqref{wcs} in Fig.~\ref{figrhonumcsl}.

\begin{figure}[t!]
\centering
\includegraphics[width=8.2cm,trim={0.6cm 1cm 0 0},clip]{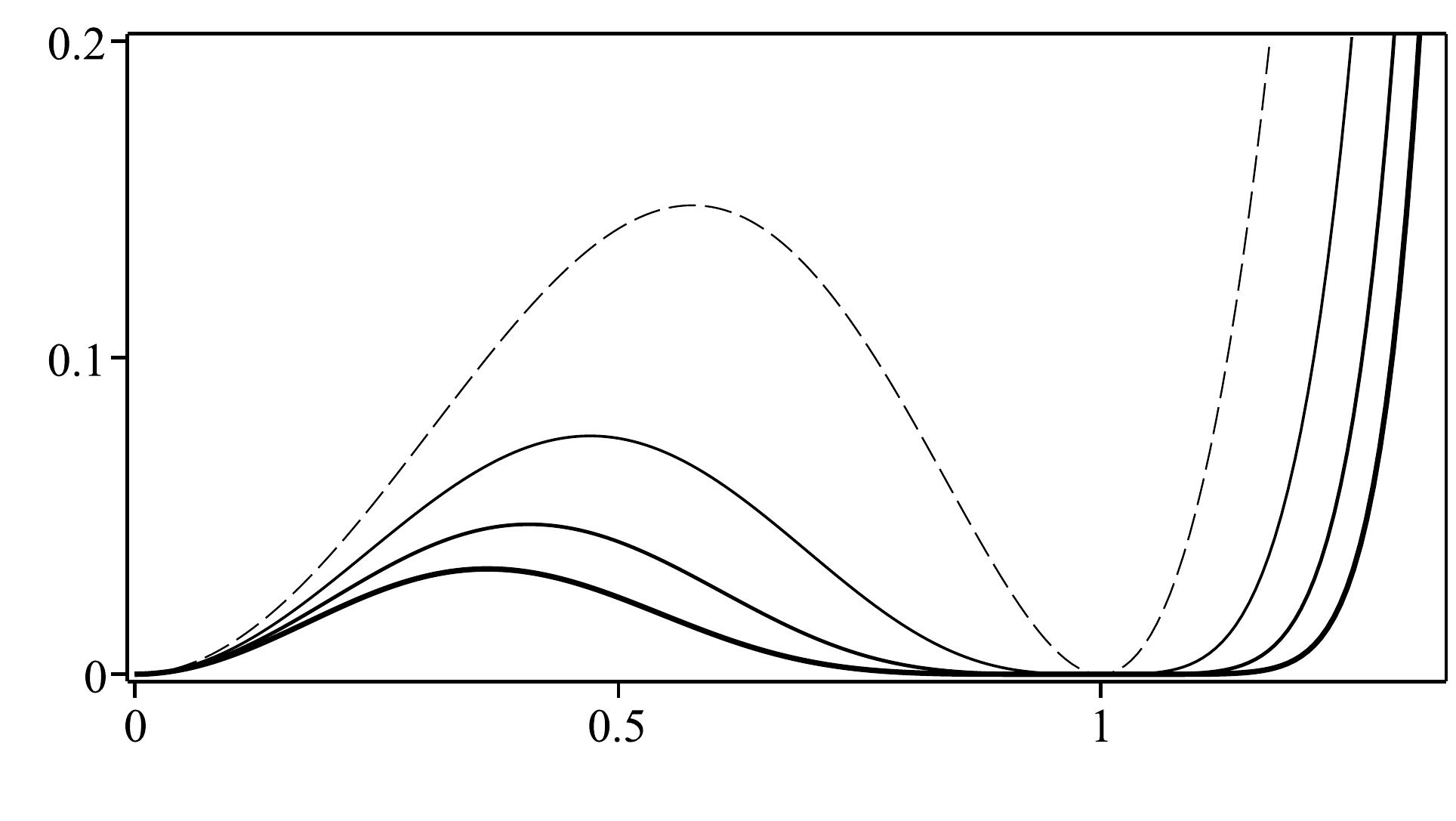}
\caption{The potential $V(g)$ in Eq.~\eqref{potnumcs} for $l=1,2,3$ and $4$. The dashed line represents the case $l=1$ and the thickness of the lines increases with $l$.}
\label{figpotnumcs}
\end{figure}

We now deal with the special case, $\alpha=0$. In this case, we get from Eqs.~\eqref{kcsc} and \eqref{potcsc} that
\bes
\bal
K(g) &= \sqrt{\frac{l+1}{2}}\, \left|1-g^2\right|^{\frac{l-1}{2}},\\ \label{potnumcs}
V(g) &= \sqrt{\frac{2}{l+1}}\,g^2\left|1-g^2\right|^{\frac{3l+1}{2}}.
\eal
\ees
The above potential is displayed in Fig.~\ref{figpotnumcs} for some values of $l$. One can show that $d^mV/dg^m|_{g=1}=0$ for $m=0,\ldots,\left\lceil{(3l-1)/2}\right\rceil$. We note here that $l=1$ recovers the standard Chern-Simons model \cite{jackiw,coreanos}, which arises for $K(g)=1$ and $V(g) = g^2\left(1-g^2\right)^2$ and engender solutions with exponential tails in a similar form of Eq.~\eqref{asystdm}.

To calculate the energy for a general $l$, one can show the auxiliar function $W(a,g)$ in Eq.~\eqref{wf} has the form
\be
W(a,g) = -\sqrt{\frac{2}{l+1}}\,a\left(1-g^2\right)^{\frac{l+1}{2}}.
\ee
The energy of the solutions in this scenario is given by Eq.~\eqref{energycs}, which leads to $E=2\pi\sqrt{2/(l+1)}$. Notice this result is different from the energy obtained in the Maxwell-Higgs model, which is constant, as one can find below Eq.~\eqref{rhow}.

\begin{figure}[t!]
\centering
\includegraphics[width=8.2cm,trim={0.6cm 1cm 0 0},clip]{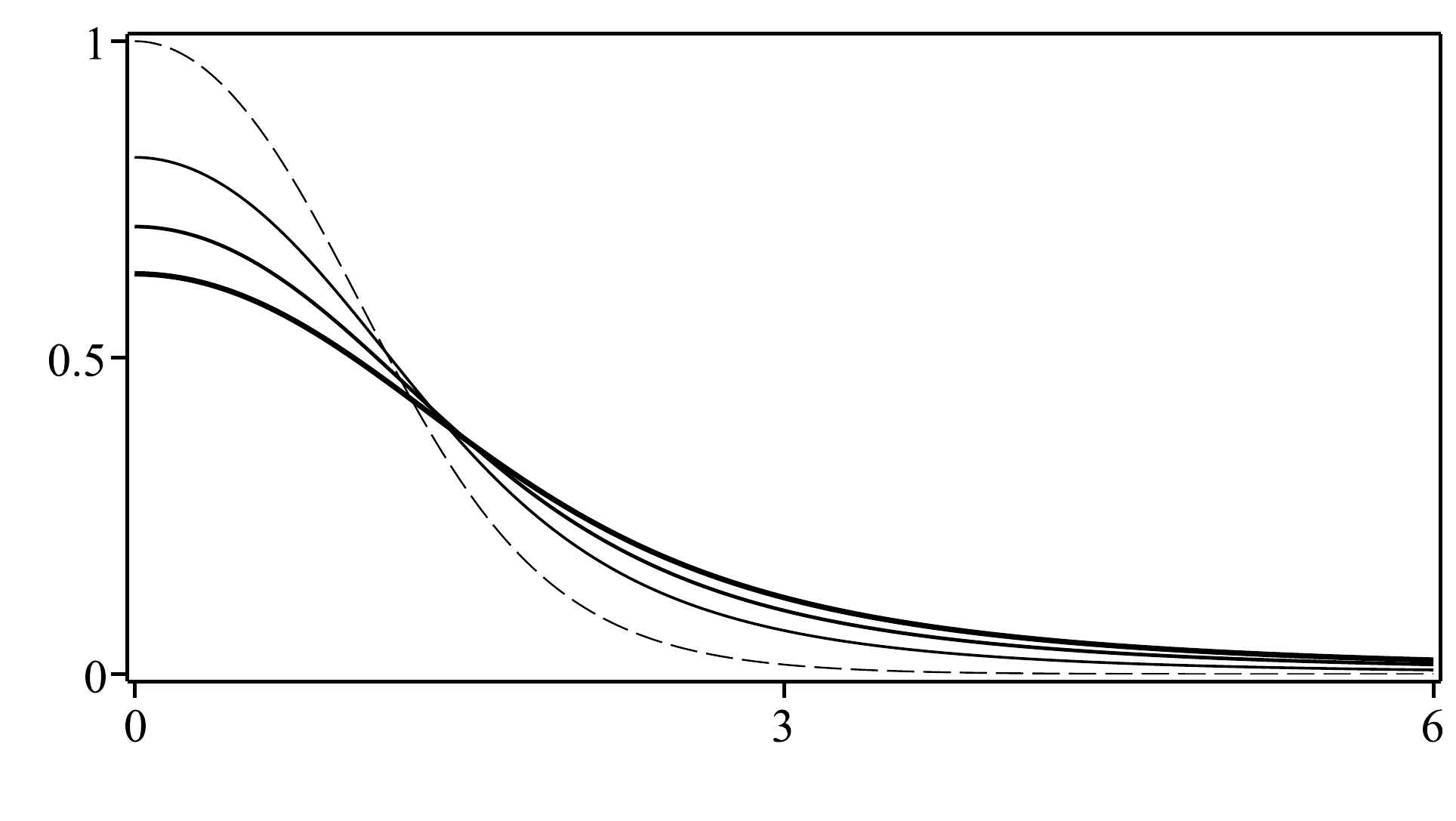}
\includegraphics[width=8.2cm,trim={0.6cm 1cm 0 0},clip]{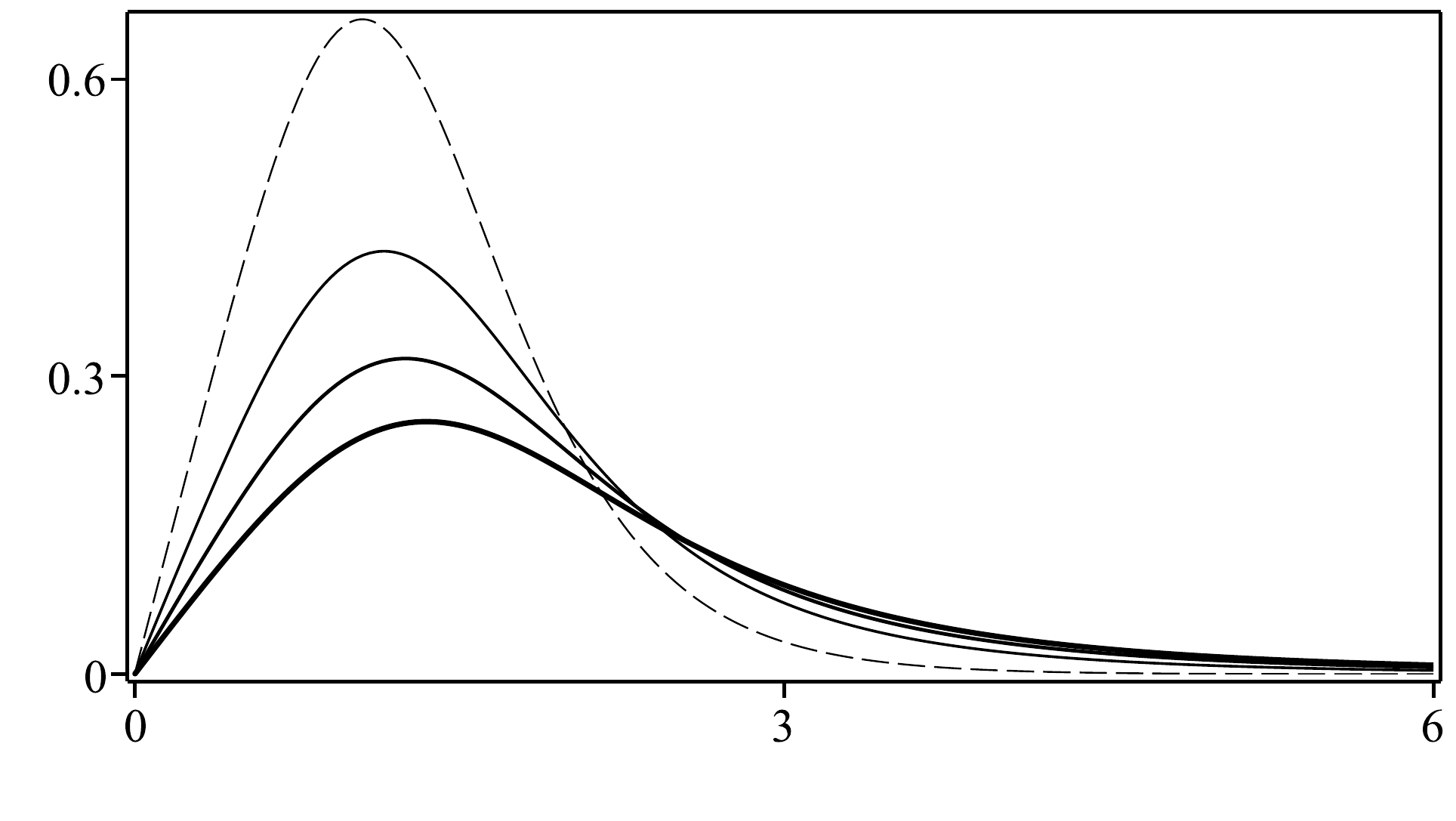}
\caption{The temporal component of the gauge field $A_0(r)$ (top) and the intensity of the electric field $|\textbf{E}(r)|$ (bottom) associated to the Chern-Simons model described by the potential in Eq.~\eqref{potnumcs} for $l=1,2,3$ and $4$. The dashed lines represent the case $l=1$ and the thickness of the lines increases with $l$.}
\label{figenumcs}
\end{figure}

\begin{figure}[t!]
\centering
\includegraphics[width=8.2cm,trim={0.6cm 1cm 0 0},clip]{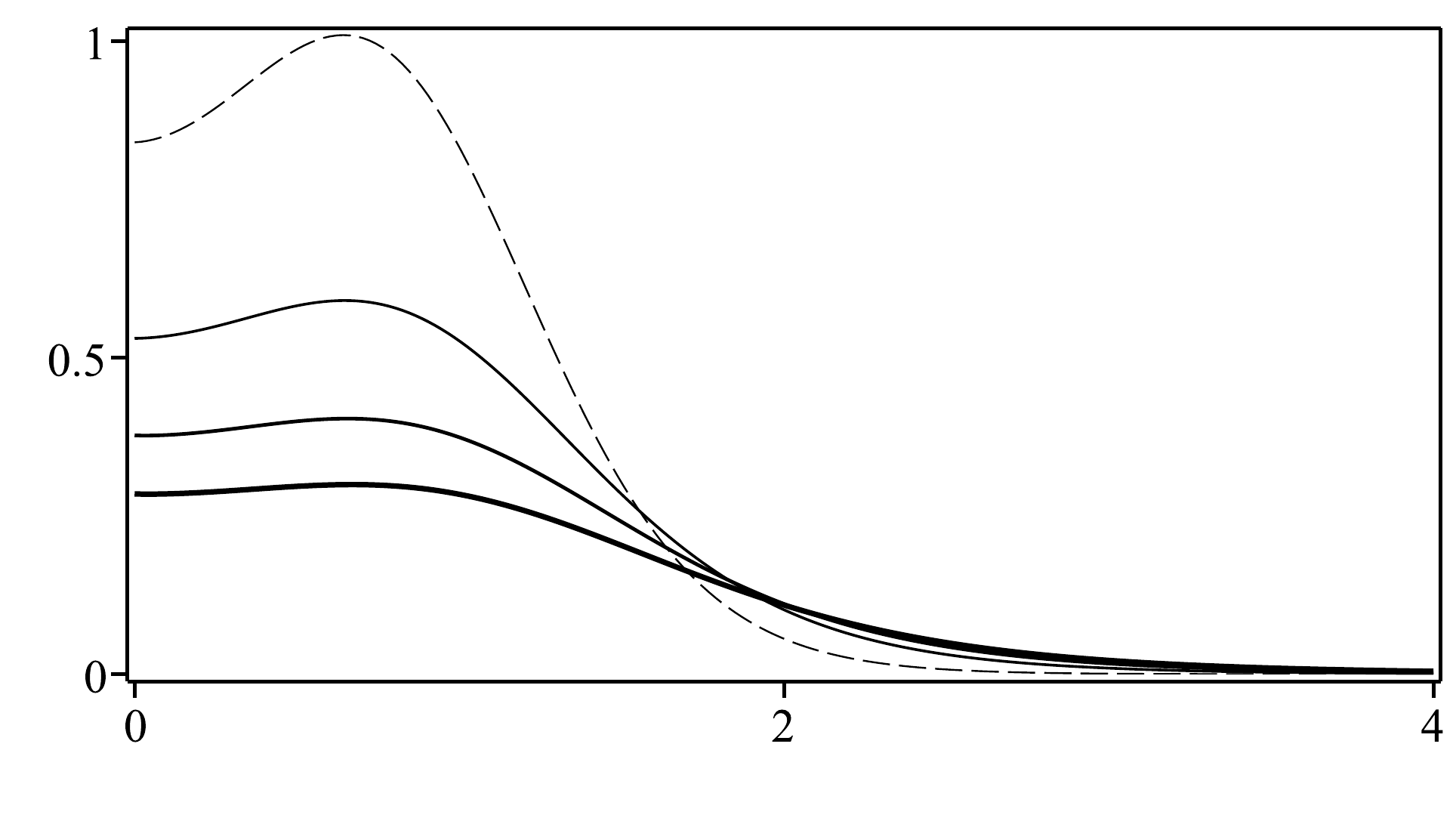}
\caption{The energy density $\rho(r)$ associated to the Chern-Simons model described by the potential in Eq.~\eqref{potnumcs} for $l=1,2,3$ and $4$. The dashed lines represent the case $l=1$ and the thickness of the lines increases with $l$.}
\label{figerhonumcs}
\end{figure}

As stated before, even though the solutions and magnetic field are the same of Fig.~\ref{figsolbnum}, here we have novel features. The function $A_0$ can be calculated from Eq.~\eqref{a0csansatz} and the intensity of the electric field $|\textbf{E}|=|A_0^\prime|$ from \eqref{ecsansatz}. The energy density can be calculated from Eq.~\eqref{wcs}. Since we only know the numerical solutions, we estimate the asymptotic behavior of these quantities using the results for the tail of $a(r)$ and $g(r)$ in Eq.~\eqref{solasynum}:
\bes
\bal
A_0(r) &\approx \sqrt{\frac{2}{l+1}}\left(l-1\right)^{-\frac{l+1}{l-1}}\,r^{-\frac{l+1}{l-1}}\\
|\textbf{E}(r)| &\approx \sqrt{2(l+1)}\left(l-1\right)^{-\frac{2l}{l-1}}\,r^{-\frac{2l}{l-1}}\\
\rho(r) &\approx \sqrt{\frac{2}{l+1}}\,(l+3)\left(l-1\right)^{-\frac{3l+1}{l-1}}\,r^{-\frac{3l+1}{l-1}}.
\eal
\ees
So, as $l$ increases, these quantities get a larger tail, which shows the long range behavior of the vortex. We then use the numerical solutions of \eqref{eqnumm} and plot $A_0$ and the intensity of the electric field $|\textbf{E}|$ in Fig.~\ref{figenumcs}. The energy density from Eq.~\eqref{wcs} is shown in Fig.~\ref{figerhonumcs}. Notice that the behavior of this case $(\alpha=0)$ is different from the one in Eq.~\eqref{a0erhocsc}. The tail of the aforementioned quantities is larger here, as the powers of $r$ are smaller in the case $\alpha>0$.

\section{Conclusion}\label{E}

In this work, we have investigated the presence of vortices with a long range behavior in Maxwell-Higgs and Chern-Simons-Higgs models. We have used the formalism developed in Ref.~\cite{godvortex} to find a manner to calculate the energy without knowing the explicit solutions and first order equations that are compatible with the equations of motion that dictate the form of the field.

Considering a pair of solutions $a(r)$ and $g(r)$, associated to a vortex with magnetic field $B(r)$ that obey a specific class of first order differential equations, we have developed a method to find Maxwell and Chern-Simons models that support them. This allows us to make a connection between the two aforementioned scenarios. One must be careful, though, since the Chern-Simons model brings an extra degree of freedom, the electric field, when compared to the Maxwell case.

By using the above procedure, we have found novel vortex configurations that support polynomial tails. As one knows, the standard vortex considered in each scenario in Refs.~\cite{NO,jackiw,coreanos} engenders a tail that dies out exponentially. Since our vortices go slower to their boundary conditions, we called them long range vortices. The presence of long range vortices has specific interest: they describe localized excitations that attain distinct collective behavior, when compared to standard vortices. In this sense, they lead to scenarios that are different from the standard situation, and may foster the study of long range vortices in the case of nonrelativistic systems like the Bose-Einstein condensates, which are known to support vortex excitations. 
Another issue of interest concerns the problem examined in Ref. \cite{jhep}, connecting conformal quantum mechanics models and equations of the KdV hierarchy. It suggests to inquire about the possibility to relate vortices with long range tails to models that admit analytic solutions in the form of vortices with exponentially dying tails. Moreover, the above results motivate us to investigate other systems, with relativistic or nonrelativistic matter, to find new systems and solutions that engender the novel long range behavior that we have found in the present work. In the nonrelativistic case, in the case of Bose-Einstein condensates with Rydberg atoms, for instance, one knows that atoms with very large principal quantum number engender long range interactions and relatively long lifetimes, and this can be used to process quantum information and may induce the presence of vortices with long range tails. Since the experimental and theoretical studies are now bringing these possibilities into play, the search for models that support long range excitations is a topic of current interest \cite{Ry,Detect,R1,R2,R3,R4,R5,R6,R7}.

\acknowledgements{The work is supported by the Brazilian agencies Coordena\c{c}\~ao de Aperfei\c{c}oamento de Pessoal de N\'ivel Superior (CAPES), grant No.~88887.463746/2019-00 (MAM), Conselho Nacional de Desenvolvimento Cient\'ifico e Tecnol\'ogico (CNPq), grants Nos. 140490/2018-3 (IA), 306614/2014-6 (DB), 404913/2018-0 (DB) and 306504/2018-9 (RM), and by Paraiba State Research Foundation (FAPESQ-PB) grants Nos. 0003/2019 (RM) and 0015/2019 (DB).}

\end{document}